\documentclass[twocolumn]{revtex4-2}
\usepackage[dvipdfmx]{graphicx}
\usepackage{amsmath,amsbsy,amssymb}
\usepackage{bm}
\usepackage{mathrsfs}
\usepackage{textgreek}
\usepackage{mathrsfs}
\usepackage{mleftright}
\usepackage{braket}
\usepackage{physics}
\usepackage{hyperref}

\usepackage{empheq}

\graphicspath{{./figures/}}

\usepackage{color}

\newcommand{\diff}{\mathrm{d}}

\newcommand{\imu}{\mathrm{i}}
\newcommand{\epn}{\mathrm{e}}

\newcommand{\ua}{\uparrow}
\newcommand{\da}{\downarrow}
\newcommand{\dg}{\dagger}
\newcommand{\la}{\langle}
\newcommand{\ra}{\rangle}

\newcommand{\sg}{\sigma}
\newcommand{\gm}{\gamma}
\newcommand{\ep}{\varepsilon}

\begin{document}

\title{
Superconductivity-enhanced phonon angular momentum
}

\author{
Natsuki Okada$^1$, Philipp Werner$^2$, and Shintaro Hoshino$^1$}

\affiliation{
$^1$Department of Physics, Chiba University, Chiba 263-8522, Japan
\\
$^2$Department of Physics, University of Fribourg, 1700 Fribourg, Switzerland
}

\date{\today}

\begin{abstract}

We theoretically investigate the properties of phonon angular momentum in the superconducting state, using fulleride compounds in an external magnetic field as a model system. 
The electron orbital angular momentum injected by an external field is transferred to the phonon subsystem via electron--phonon coupling. 
We show that this field-induced phonon angular momentum is significantly enhanced and undergoes a sign reversal upon entering the superconducting state. 
In the normal state, the dominant energy scale governing the response function is the electronic bandwidth $D$. In the superconducting state, the phonon energy scale $\omega_1$ enters the denominator, leading to an enhancement of order $D/\omega_1$. The observed sign change in the response can be explained by the competition between Fermi surface and Fermi volume contributions. 
\end{abstract}

\maketitle

\paragraph*{Introduction.---}

In recent years, phonon angular momentum (PAM) has attracted considerable attention as an exploitable degree of freedom associated with lattice vibrations \cite{Zhang14, Juraschek25}. Both theoretical and experimental studies demonstrated that PAM can be generated and controlled via external driving. For example, terahertz light can induce circularly polarized phonon motion and thereby generate nonzero PAM. Such a driven phonon system can transfer angular momentum to the electronic sector, inducing electronic angular momentum~\cite{Klebl25,Yao25,Sato25}. In addition, thermal transport can give rise to the phonon angular momentum Hall effect, which provides another route to manipulate PAM \cite{Park20}. 
The application of an external magnetic field to either the electronic or phononic degrees of freedom leads to a phonon Zeeman effect \cite{Juraschek17}. 
These developments establish PAM as a relevant quantity that couples to magnetization and spin, and demonstrate that it can be actively controlled by external perturbations.

Despite this progress, the impact of superconductivity on PAM remains unexplored. 
Since superconductivity fundamentally modifies low-energy electronic excitations through the formation of an energy gap, it is natural to expect nontrivial changes in PAM across the superconducting transition. 
Clarifying this interplay is essential for understanding the behavior of angular momentum in coupled electron--phonon systems. 
Here, we present a microscopic theory which elucidates how the superconducting state affects PAM in fulleride compounds.

Alkali-doped fullerides exhibit $s$-wave superconductivity with nearly isotropic gap, and with transition temperatures up to approximately $35$~K~\cite{Gunnarsson97,Capone09,Nomura16,Takabayashi16}. Their pairing mechanism is nontrivial, since intramolecular Jahn--Teller phonons play an essential role. In A$_3$C$_{60}$ compounds, the threefold-degenerate lowest unoccupied molecular orbitals (LUMO) are coupled to fivefold-degenerate Jahn--Teller phonon modes. This coupling effectively reduces the Hund's coupling, favoring low-spin configurations and stabilizing intraorbital spin-singlet Cooper pairs. Since the molecular vibrations are dominantly local, fullerides provide a minimal and controlled platform for investigating PAM in the superconducting state.

In this work, we analyze the temperature dependence of PAM induced by an orbital Zeeman field by solving the Eliashberg equations for a Jahn--Teller Hubbard model relevant to fulleride compounds. 
We find that the PAM, which is positive in the metallic state, changes sign  
below the superconducting transition temperature, with its magnitude 
significantly 
increasing upon cooling. 
We clarify the microscopic origin of this superconductivity-driven sign reversal and discuss the possibility of controlling PAM across the superconducting transition

\paragraph*{Model.---}

We study a three-orbital Hubbard model coupled to local intramolecular phonons with both isotropic and Jahn--Teller anisotropic modes~\cite{Han00,Nomura15_2,Kaga22,Okada26}. 
The analysis is based on 
the spherical tensor formalism~\cite{Iachello_book,Judd67,Ceulemans94,Takahashi26}, which enables a systematic
classification of electronic and phononic degrees of freedom based on their angular momenta. 
The electronic degrees of freedom are described by $p$ fermions ($\ell=1$), corresponding to the threefold-degenerate $t_{1u}$ molecular orbitals, whose character is analogous to atomic $p$ orbitals. 
The corresponding annihilation operator is denoted by $p_{im\sg}$ at site $i$, with magnetic quantum number $m=-1,0,1$ and spin $\sg=\ua,\da$ \cite{suppl,Takahashi26}.
To construct the Hamiltonian, we introduce the 
quadrupole (orbital) tensors as
\begin{align}
    \mathscr Q_{iM} &= - 2\, \qty[p_i^\dg \otimes \tilde p_i]_M^{(2)} .
\end{align}
Here, the orbital tensor product is defined as
$    [a\otimes~b]_{M}^{(L)} = \sum_{mm'} C^{LM}_{\ell m\ell m'} \sum_{\sg} a_{m\sg} b_{m'\sg}$, where $a_m$ and $b_m$ are rank-$\ell$ tensors and $C$ denotes the Clebsch--Gordan coefficients~\cite{Varshalovich_book}.
The operators 
$p_{im\sg}^\dg$ and $\tilde p_{im\sg} = (-1)^{1-m} p_{i,-m,\sg}$ transform as rank-$L=1$ tensor operators.

The total Hamiltonian is decomposed as
$\mathscr{H}=\mathscr{H}_e+\mathscr{H}_p+\mathscr H_{ep}$.
The electronic part is given by
\begin{align}
    \mathscr H_e &= \mathscr H_{0e}  + \sum_i \qty( \frac { \tilde U} 2 N_i^2  + \frac J 2 \mathscr Q_i\cdot \mathscr Q_i -\mu N_i ) ,
\end{align}
where $N_i$ is the number operator and $\tilde U=U -\tfrac 3 4J$ is the effective Coulomb interaction parameter. 
The chemical potential $\mu$ controls the average electron filling, which we fix to $\la N_i \ra = 3$. 
The scalar product for bosonic tensors is defined as $A\cdot B = \sum_M (-1)^M A_M B_{-M}$.
$\mathscr H_{0e}$ describes the kinetic energy arising from intersite electron hopping.

The electron--phonon interaction is given by
\begin{align}
\mathscr H_{ep} &= \sum_i \big( g_0 \, 
\delta \! N_i 
\, s_i^\dg + g_1 \mathscr Q_i\cdot d_i^\dg \big)  + {\rm H.c.} ,
\end{align}
where $s_i^\dg$ is the creation operator for the nondegenerate $A_g$ mode ($\ell=0$, the $s$ boson), $d_{iM}^\dg$ corresponds to the fivefold-degenerate $H_g$ modes ($\ell=2$, the $d$ boson), and we introduced the charge fluctuation operator $\delta \! N_i = N_i-\la N_i \ra$.
Finally, the phonon Hamiltonian is given by $\mathscr H_p = \sum_i(\omega_0 s_i^\dg s_i + \omega_1 d_i^\dg \cdot \tilde d_i)$ with $\tilde d_{iM} = (-1)^M d_{i,-M}$, where the Planck constant is set to unity ($\hbar = 1$). 
In this formalism, we assume rotational invariance of the local interactions, which is approximately satisfied in fulleride compounds~\cite{Nomura15}.

\paragraph*{Angular momentum operators.---}

To induce phonon angular momentum (PAM), we consider an external magnetic field. 
Let us first introduce the orbital angular momentum (OAM) of the electrons, i.e., $p$ fermions, and the PAM of the Jahn--Teller phonons, i.e., $d$ bosons, as
\begin{align}
    L_{i,\mathrm{el}}^z
    &=
    \sum_{m=-1}^1 m \sum_\sg p_{im\sg}^\dg p_{im\sg}
    = -2 \qty[p_i^\dg \otimes \tilde p_i]_0^{(1)},
    \\
    L_{i,\mathrm{ph}}^z 
    &=
    \sum_{M=-2}^{2} M d_{iM}^\dg d_{iM}
    = \sqrt{10} \qty[d_i^\dg \otimes \tilde d_i]_0^{(1)}.
    \label{eq:Lph_angular_basis}
\end{align}
When an external magnetic field is applied along the $z$ direction, its coupling to the system is described by the Zeeman term
\begin{align}
    \mathscr{H}_{\rm ext}
    = \mu_{\rm B} B \sum_i \qty( \mathrm g_L L_{i,\rm el}^z + \mathrm g_S\, S_{i,\rm el}^z - \mathrm g_{\rm ph}\, L_{i,\rm ph}^z ),
    \label{eq:ext_field}
\end{align}
where the spin angular momentum (SAM) operator is defined as $S^z_{i,\rm el} = \frac 1 2\sum_{m} (p_{im\ua}^\dg p_{im\ua} - p_{im\da}^\dg p_{im\da})$. 
Here, $\mu_{\rm B} = \hbar |e| / 2m_e$ is the Bohr magneton. 
The relative signs reflect the opposite charges of electrons and ions, while $\mathrm g_L=O(1)$~\cite{Tosatti96} and $\mathrm g_S=2$ denote the orbital and spin $\mathrm g$-factors, respectively. In this paper we choose $\mathrm g_L = 1$. 
The phonon $\mathrm g$-factor, which is defined through Eq.~\eqref{eq:ext_field}, is suppressed by the inverse ionic mass ($\mathrm g_{\rm ph}\sim m_{\rm el}/M_{\rm ion}$) and is therefore much smaller than its electronic counterparts. 
Accordingly, we neglect this contribution in the present calculations, 
but retain its sign in the analysis of response functions. 
In the following, we assume translational invariance and omit the site index $i$ for simplicity.

\paragraph*{Method of solution.---
}

The model defined above is solved by combining dynamical mean-field theory with a perturbative treatment of the interactions, i.e., the local Eliashberg framework~\cite{Kaga22,Ishida25,Okada26}. 
This approach is expected to provide a reliable approximation for three-dimensional systems in the weak-coupling regime.
For simplicity, the noninteracting electronic density of states is assumed to have a featureless semielliptical form:
    $\rho(\ep) = 
    \frac{2}{\pi D^2}\sqrt{D^2 - \ep^2}$,
where the bandwidth is $2D$. 
Such a structureless density of states is suitable for isolating the effects of electron--phonon coupling, allowing us to extract generic behaviors that are not tied to specific details of the underlying band structure.
In the numerical calculation, we choose the parameters $\omega_0=0.318$,
$\omega_1=0.217$,
$g_0=0.14$,
$g_1=0.073$,  
$U=0.2$ and $J=0.031$ as in Ref.~\cite{Okada26}.

\paragraph*{Temperature dependence.---}

\begin{figure}[tb]
  \centering
  \includegraphics[width=1.0\linewidth]{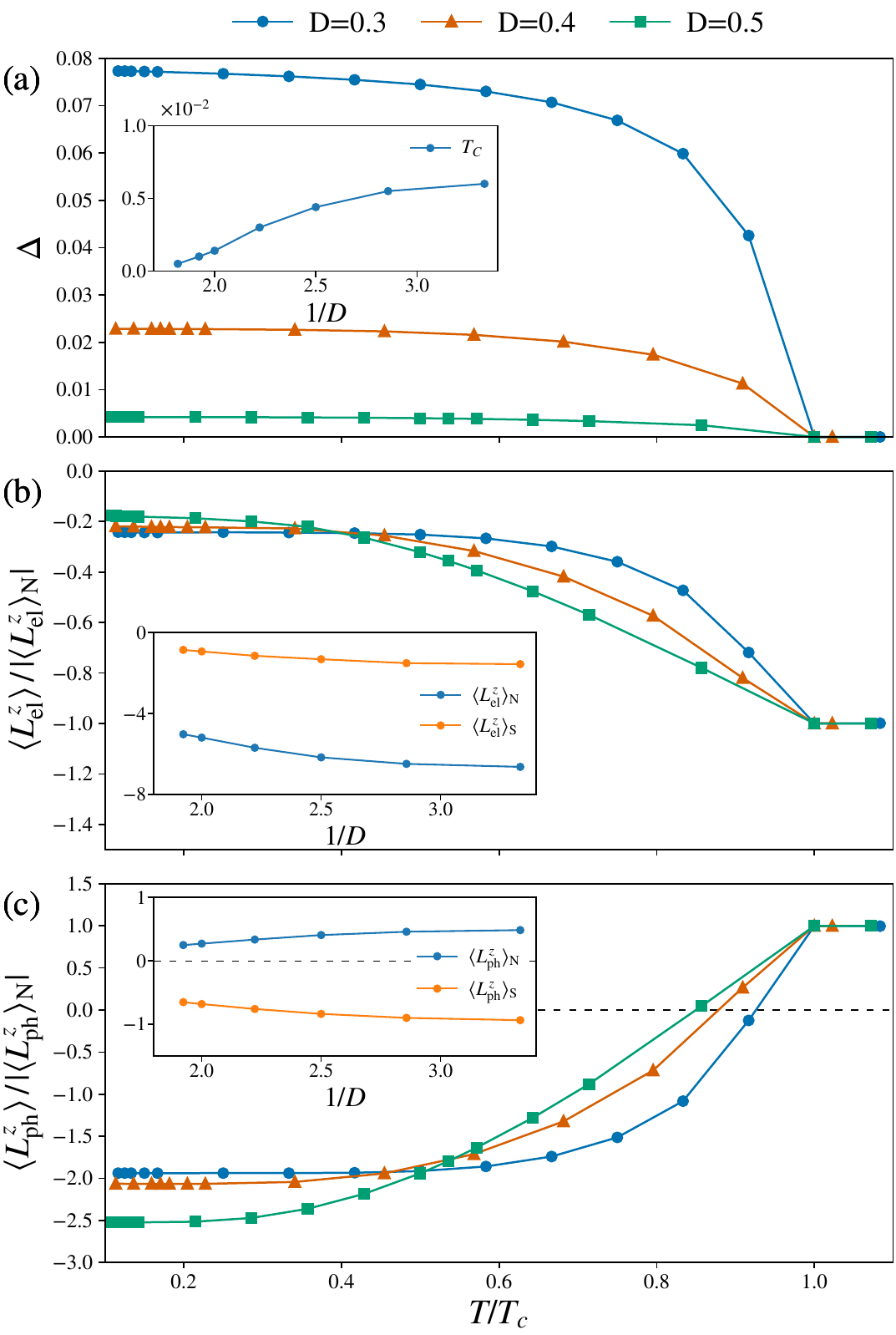}
  \caption{
Temperature dependence of (a) the superconducting gap function [inset: bandwidth dependence of the transition temperature] and (b,c) the electronic and phononic orbital angular momenta. 
In (b) and (c), the vertical axis is normalized by the magnitude of the corresponding quantity in the normal state, while the magnitudes are shown in the insets.
  }
  \label{fig:temperature_dep}
\end{figure}

To identify the superconducting regime, we first show in Fig.~\ref{fig:temperature_dep}(a) the temperature dependence of the gap function of the spin-singlet pairs at the lowest Matsubara frequency, $\Delta \equiv \Delta(\imu\omega_{n=0})$, for several values of the bandwidth $D$. 
Here, the temperature is normalized by the superconducting transition temperature $T_c$ for each $D$. 
For all bandwidths, the gap continuously decreases toward $T_c$, exhibiting a typical second-order phase transition. 
The inset shows the bandwidth dependence of $T_c$: $T_c$ increases as $1/D$ increases, i.e., as the bandwidth $D$ decreases. 
The ratio $\Delta/T_c$ is enhanced compared to the BCS value, reflecting the strong-coupling nature with retardation effects~\cite{Scalapino_book, Kaga22}.

Next, we consider the effect of a weak magnetic field ($h \equiv \mathrm g_L \mu_{\rm B} B = 0.001$). 
The resulting temperature dependence of the electronic orbital angular momentum (OAM) is shown in Fig.~\ref{fig:temperature_dep}(b), normalized by its absolute value in the metallic state just above $T_c$.
Note that the electronic angular momentum must be negative as derived from Eq.~\eqref{eq:ext_field}.
In the normal state, the OAM is nearly temperature independent. 
Below $T_c$, its magnitude decreases as the superconducting order develops and it eventually saturates to a nonzero value at low temperatures. 
This implies that the orbital response to the Zeeman field is not completely suppressed in the superconducting state, as detailed in Ref.~\cite{Okada26}. 
The magnitude of the OAM in the superconducting state also decreases with increasing bandwidth $D$. This  
behavior can be attributed to the reduced retardation effects as the system approaches the BCS limit for larger $D$, leading to a suppressed  orbital response.

We now turn to the central result of this work: the temperature dependence of the phonon angular momentum (PAM), shown in Fig.~\ref{fig:temperature_dep}(c). 
In the normal state, the PAM takes a positive value, but it changes sign 
below
the superconducting transition. 
Upon further cooling, the magnitude of the PAM increases relative to its normal-state value, with a more pronounced enhancement for larger bandwidths. 
This behavior suggests that the contributions from normal-state electrons and superconducting Cooper pairs have opposite signs.

The inset of Fig.~\ref{fig:temperature_dep}(c) shows the $1/D$ dependence of the PAM in the normal state just above $T_c$ and in the superconducting state at sufficiently low temperature. 
In both phases, the magnitude of the PAM increases as the bandwidth $D$ decreases. 
The $D$-dependence is however more pronounced in the normal state, so that the ratio $\la L_{\rm ph}^z\ra _{\rm S}/\la L_{\rm ph}^z\ra_{\rm N}$ becomes larger with increasing bandwidth $D$.

\paragraph*{Phonon self-energy.---}

\begin{figure}[tb]
  \centering
  \includegraphics[width=1.0\linewidth]{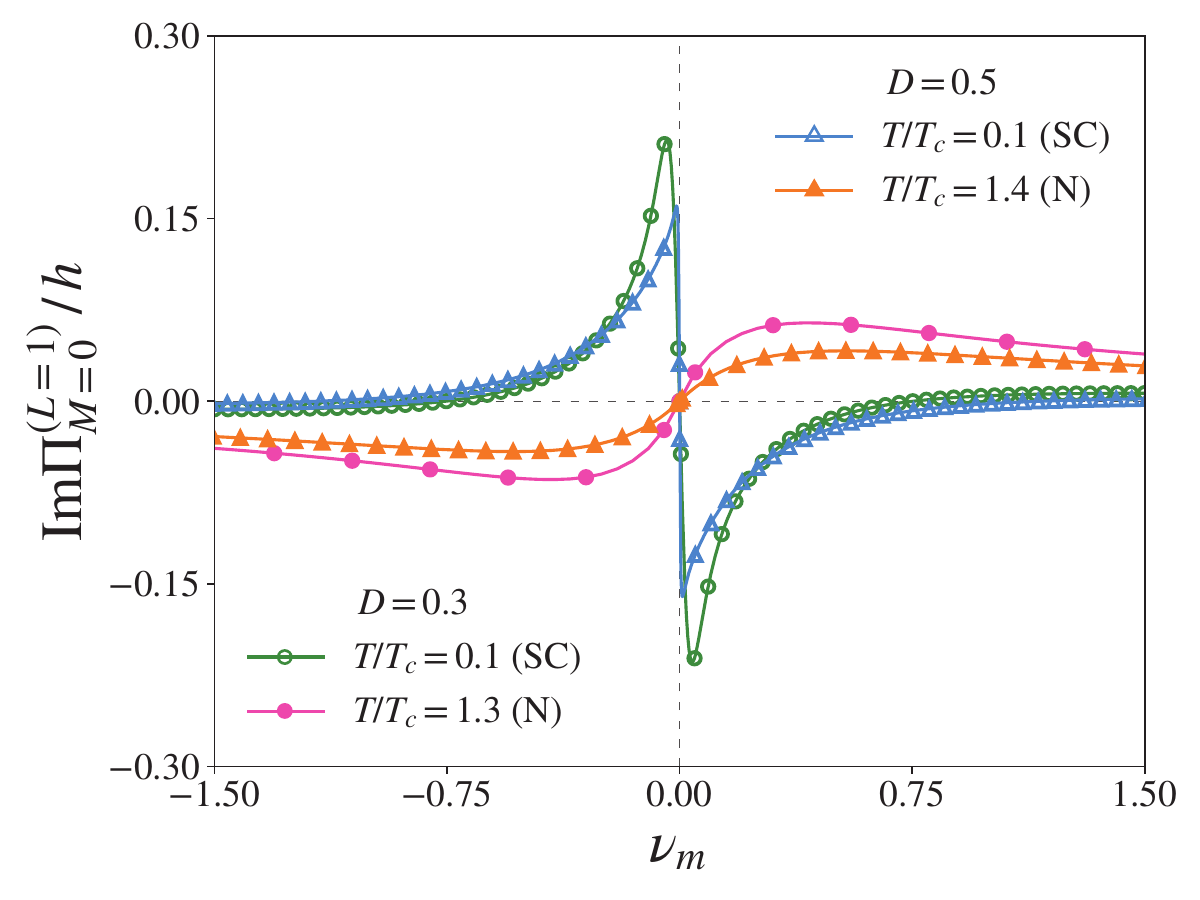}
  \caption{
Field-induced phonon self-energy in a weak magnetic field. 
Circles and triangles correspond to $D=0.3$ and $D=0.5$, while filled and open symbols indicate the normal and superconducting states.
}
  \label{fig:selfenergy}
\end{figure}

To elucidate the origin of the change in PAM across the superconducting transition, we analyze the phonon self-energy, 
which  incorporates the effect of the orbital magnetic field. 
The relevant 
contribution
is given by the quadrupole correlation 
\begin{align}
        \Pi_M^{(L=1)}(\tau) = g_1^2 \ \Big\la 
        \big[\mathscr Q(\tau) \otimes \mathscr Q \big]_M^{(1)}
        \Big\ra,
        \label{eq:phonon_self_L1}
\end{align}
which is induced by the
magnetic field $h=\mathrm g_L\mu_{\rm B}B$.
The imaginary-time dependence of operators is defined by $A(\tau) = \epn^{\tau \mathscr H} A \epn^{-\tau \mathscr H}$.
Figure~\ref{fig:selfenergy} shows the $L=1$ component 
$\mathrm{Im}\,\Pi_{M=0}^{(1)}(\imu\nu_m)$ in Fourier space. 
Here, we consider the bandwidths $D=0.3$ and $D=0.5$, and compare the results in the normal and superconducting states.
The field-induced self-energy in Fig.~\ref{fig:selfenergy} is an odd function of frequency, whose low-energy behavior is characterized by $\Pi_{M=0}^{(1)}(\imu \nu) \propto \Gamma \imu \nu$. 
We find that $\Gamma$ 
is positive in the normal state, whereas it changes sign and is strongly enhanced in magnitude in the superconducting state. 
Furthermore, the frequency range over which this linear behavior is observed is of the order of the bandwidth ($\sim D$) in the normal state, while it is reduced to the scale of the superconducting gap ($\sim \Delta$) in the superconducting state.

Let us now consider the physical meaning of the coefficient $\Gamma$. 
By integrating out the electronic degrees of freedom, one obtains an effective action for the phonons, from which effective classical equations of motion can be derived. 
In particular, by constructing the $zx$- and $yz$-type displacement operators, $\phi_{zx}$ and $\phi_{yz}$ defined from the $d$ bosons, one finds that they satisfy the following equations in the low-frequency regime below $\Delta$:
\begin{equation}
\begin{aligned}
    \ddot \phi_{zx} &= - \tilde\omega_1^2\phi_{zx} - 2\omega_1 \Gamma \dot \phi_{yz},
    \\
    \ddot \phi_{yz} &= - \tilde\omega_1^2\phi_{yz} + 2\omega_1 \Gamma \dot \phi_{zx},
    \label{eq:eff_interp}
\end{aligned}
\end{equation}
(see Ref.~\cite{suppl} for the derivation). 
Here, $\tilde \omega_1$ denotes the renormalized phonon frequency including electron--phonon interactions. 
The second terms on the right-hand side have the same structure as the Lorentz force acting on a charged particle with charge $q$ and mass $M$, whose equation of motion is given by $M\ddot {\bm r} = q \dot {\bm r}\times \bm B$.

\paragraph*{Low-energy effective theory.---}

We now discuss how the sign and magnitude of the PAM are determined. 
In the superconducting state, 
 the low-frequency behavior can be obtained by expanding the Fourier transform of Eq.~\eqref{eq:phonon_self_L1} with respect to the bosonic Matsubara frequency $\nu_m$. 
The characteristic feature of the phonon self-energy can be captured in the BCS limit (i.e., no retardation effect), and the result is analytically obtained as \cite{suppl}
\begin{align}
    \Pi_{M=0}^{(1)}(\imu\nu_m)&\simeq - \sqrt{\frac{18}{5}}\frac{2 h g_1^2}{D^2 \Delta}  \imu \nu_m \equiv \sqrt{\frac{18}{5}} \Gamma_{\rm S} \,  \imu \nu_m ,
    \label{eq:SC_Gamma}
\end{align}
where the denominator contains the small energy scale $\Delta$ in the superconducting state.
Accordingly, based on Eq.~\eqref{eq:eff_interp}, we obtain the following expression for the effective magnetic field $B_{\rm eff}$ acting on the phonons, 
\begin{align}
    \frac{qB_{\rm eff}}{M_{\rm ion}}
    = 2\omega_1 \Gamma_{\rm S} 
    \sim - 
    \  \qty(\frac{\omega_1}{D})^2 
    \  \frac{\lambda_1}{\Delta}
    \  \frac{e B}{m_{\rm el}} .
\end{align}
Here, $\lambda_1 = 3g_1^2/4\omega_1 \simeq 0.033~\mathrm{eV}$ denotes the phonon-mediated interaction. 
Due to the large mass ratio $M_{\rm ion}/m_{\rm el} \sim 10^4$, the effective magnetic field acting on the phonons is greatly enhanced. 
This enhancement is further amplified in the superconducting state, where the denominator involves the gap $\Delta$, in contrast to the normal state, where it is replaced by the bandwidth $D$ ($\gg \Delta$), as shown below.

Next, we examine the phonon self-energy in the normal state. 
At low temperatures and low bosonic frequencies, the phonon self-energy takes the form $\Pi_{0}^{(1)}(\imu \nu_m) \simeq \sqrt{\frac{18}{5}}\Gamma_{\rm N} \imu\nu_m$, with \cite{suppl}
\begin{align}
    \Gamma_{\rm N}
    &\simeq 4 h g_1^2 \int_{-\infty}^\infty \frac{\diff \omega}{2\pi} \Big[ G'_{\rm sing}(\imu \omega)G'_{\rm reg}(\imu \omega) + \big|G'_{\rm reg}(\imu \omega)\big|^2 \Big] .
    \label{eq:Gamma_N_orig}
\end{align}
Here, we separated $G'(\imu\omega)$, the derivative  of the non-interacting local electron Green's function $G(z)$, into the regular and singular contributions.
In deriving this expression, we assume $h \ll T \ll D$. 
The singular part $G'_{\rm sing}(\imu\omega) \propto \delta(\omega)$ originates from the Fermi-surface contribution, 
and our analysis shows that its 
sign differs from that of the Fermi-volume contribution in the second term of Eq.~\eqref{eq:Gamma_N_orig}~\cite{suppl}. 
The total contribution in the normal state is then given by
\begin{align}
    \Gamma_{\rm N} &= \frac{8(\pi-2)hg_1^2}{\pi D^3} ,
\end{align}
which has the opposite sign compared to the superconducting case, due to the Fermi-surface contribution. 
The ratio of the responses is given by $\Gamma_{\rm S}/\Gamma_{\rm N} \sim - D/\Delta$. 
Since the superconducting gap is typically much smaller than the electronic bandwidth, this implies a strong enhancement of the effective Lorentz force below $T_c$.

We next consider the magnitude of the PAM, i.e., the static observable obtained after performing the bosonic frequency summation. 
As discussed in connection with Fig.~\ref{fig:selfenergy}, the $\nu_m$-linear regime is cut off at a characteristic energy scale: $\sim D$ for $T>T_c$ and $\sim \Delta$ for $T\ll T_c$. 
Taking this characteristic energy scale into account, we obtain \cite{suppl}
\begin{align}
    \frac{\la L_{\rm ph}^z\ra_{\rm S}}{\la L_{\rm ph}^z\ra_{\rm N}} 
    \ \sim \ 
    \frac{\Gamma_{\rm S}}{\Gamma_{\rm N}} \cdot \frac{\Delta}{\omega_1}
    \ \sim \ 
    - \frac{D}{\omega_1}
    \label{eq:PAM_conclusion}
\end{align}
with $\la L_z^{\rm ph} \ra_{\rm N} \simeq 8(\pi-2) g_1^2 h / D^3 $.
Notably, the superconducting gap no longer appears in the final expression and the ratio of the PAM in the superconducting and normal states is simply governed by the ratio of the electronic bandwidth to the phonon frequency. These quantities represent the characteristic energy scales of the electrons and phonons, respectively. 
Since they are typically well separated in superconductors, we generally expect a substantial enhancement of the field-induced phonon angular momentum  
below $T_c$.

The response of an observable $B$ to a perturbation conjugate to $A$ is given by the canonical correlation function
$\chi_{BA}(\mathrm{i}\nu_m) 
= \int_0^\beta \mathrm{d}\tau \, \langle A(\tau) B \rangle \, \epn^{\mathrm{i}\nu_m \tau}$,
whose spectral representation contains an energy denominator. 
Such a structure implies that the response function can be enhanced when a small characteristic energy scale appears in the denominator, provided that the corresponding matrix elements remain finite. 
This mechanism, when applied to the quadrupole correlation function with 
$A=B=\mathscr Q_M$, 
underlies the enhancement of the phonon self-energy [Eq.~\eqref{eq:phonon_self_L1}] in the presence of a superconducting gap $\Delta$, as seen in Eq.~\eqref{eq:SC_Gamma}.
It is analogous to the enhancement of nonreciprocal responses in superconductors, where the ratio of the signal in the superconducting state to that in the normal state is controlled by the ratio of the Fermi energy to the superconducting transition temperature~\cite{Wakatsuki17,Hoshino18,Itahashi22}.

However, the enhancement of the self-energy does not directly translate into the same enhancement of the static PAM with $A=L_{\rm el}^z$ and $B=L_{\rm ph}^z$; the enhancement through the self-energy is limited by the characteristic low  energy scale $\Delta$. 
As a result, the total contribution after frequency summation is not strongly sensitive to the superconducting gap itself.
Hence, the ratio $\la L_{\rm ph}^z\ra _{\rm S}/\la L_{\rm ph}^z\ra_{\rm N}$ is controlled by $D/\omega_1$ as shown in Eq.~\eqref{eq:PAM_conclusion}.
We note that, by reciprocity, the corresponding response function is identical to the electronic response under a phonon drive, represented by $A=L_{\rm ph}^z$ and $B=L_{\rm el}^z$.

\paragraph*{Magnetic-field dependence.---}

\begin{figure}[t]
  \centering
  \includegraphics[width=1.0\linewidth]{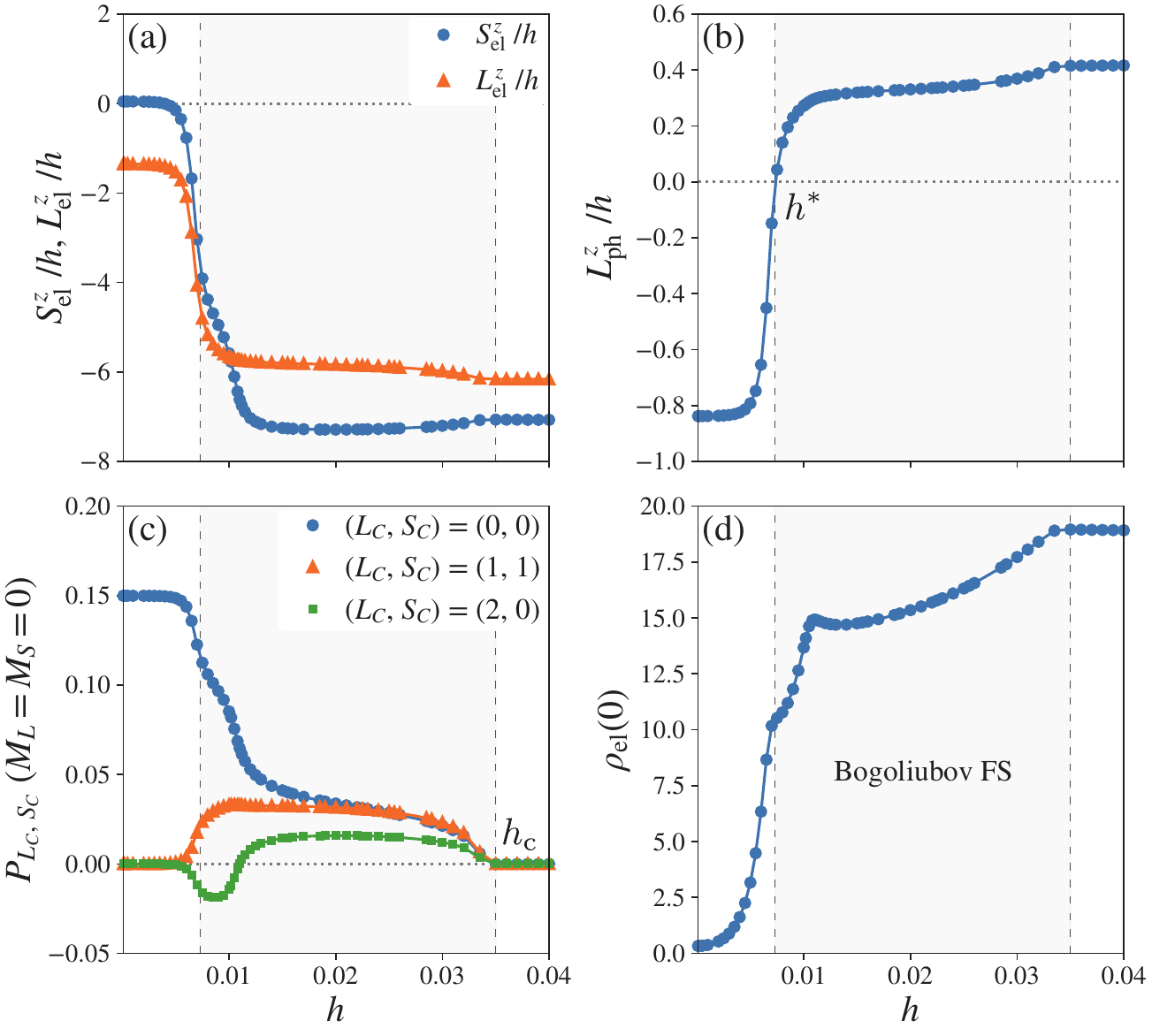}
  \caption{
Magnetic-field dependence of (a) the electronic angular momenta, OAM and SAM; 
(b) the phonon angular momentum (PAM); 
(c) the superconducting pair amplitude, characterized by the orbital and spin angular momenta of the Cooper pairs, $(L_{\rm C},S_{\rm C})$; 
and (d) the electronic spectral function at zero energy. 
The temperature and bandwidth are $T=0.001$ and $D=0.4$, respectively.
}
  \label{fig:hdep}
\end{figure}

We also discuss the magnetic-field dependence beyond the linear-response regime.
We neglect vortex states by assuming, e.g., a 
layered
system in which the magnetic field is applied parallel to the layer.
Figures~\ref{fig:hdep}(a) and (b) show the $h=g_L\mu_{\rm B}B$ dependence of the electronic and phononic angular momenta (OAM and PAM), respectively. 
For sufficiently large fields $h > h_{\rm c}\simeq 0.035$, superconductivity is destroyed [see panel (c)] and both angular momenta approach their normal-state values. 
Interestingly, however, this recovery toward the normal-state behavior occurs predominantly at a much smaller characteristic field $h^*\simeq 0.007$,
well below $h_{\rm c}$, where also the sign change in $L^z_\text{ph}$ takes place.

Figure~\ref{fig:hdep}(c) shows the superconducting order parameter, resolved into components labeled by the Cooper-pair orbital angular momentum (COAM), $L_{\rm C}$, and the Cooper-pair spin angular momentum (CSAM), $S_{\rm C}$ \cite{suppl}. 
The dominant component $(L_{\rm C},S_{\rm C})=(0,0)$ [spin-singlet, orbital-totally-symmetric] is strongly suppressed around $h\simeq h^*$ and remains nonzero, though small, for $h>h^*$, before eventually vanishing at $h=h_{\rm c}$. 
Around $h\simeq h^*$, as also shown in panel (c), subdominant components such as $(L_{\rm C},S_{\rm C})=(2,0)$ and $(L_{\rm C},S_{\rm C})=(1,1)$ are induced by the magnetic field. 
For all these components, the magnetic quantum numbers satisfy $M_L=M_S=0$, corresponding to the $z$-components, since the field is applied along the $z$ direction. 
These induced pairing components initially grow (approximately exponentially) as a function of the field, and their emergence is accompanied by a pronounced suppression of the dominant $(0,0)$ pairing amplitude.

To clarify the behavior around $h\simeq h^*$, we examine the electronic spectral function at the Fermi level. 
Figure~\ref{fig:hdep}(d) shows the magnetic-field dependence of the low-energy spectral weight $\rho_{\rm el}(0)$. 
At low fields, the spectral weight vanishes, indicating the presence of a superconducting gap. 
Around $h \sim h^*$, 
it increases rapidly, signaling the emergence of a Fermi surface. 
Since superconductivity is still present in this regime, this is naturally interpreted as a Bogoliubov Fermi surface (FS) formed by superconducting quasiparticles. 
Because Fermi-surface contributions dominate the sign of the PAM in the normal state, as discussed above, the result in Fig.~\ref{fig:hdep}(d) suggests that a similar mechanism also operates in the superconducting state for $h^*<h<h_{\rm c}$.

\paragraph*{Summay and Outlook.---}

In this work, we showed that in fulleride superconductors, a magnetic field applied to the electronic system induces a phonon angular momentum (PAM) that is strongly enhanced below the superconducting transition temperature and undergoes a sign reversal. 
The 
magnitude of the PAM is governed by the ratio of the characteristic energy scales of the electronic and phononic systems, while its sign is controlled by the presence or absence of a Fermi surface.
The latter effect is consistent with a very recent study which also hightlighted the role of the Fermi surface in electron-phonon coupled systems \cite{Chen26}.
Furthermore, we demonstrated that, in the low-energy regime, the transfer of angular momentum from the electrons to the phonons can be interpreted in terms of an effective Lorentz force acting on the phonons.

As an outlook, it would be interesting to extend this framework to systems with dispersive bosonic modes 
and other classes of materials. 
In particular, it is worth exploring the role of chiral phonons~\cite{Ishito23,Oishi24} in superconducting states, where the angular momentum is aligned with the propagation direction, as well as Kelvin modes associated with vortex lines in type-II superconductors~\cite{Barckicke26}. 
Understanding the behavior of these modes below $T_c$ is a promising direction for future work.

\paragraph*{Acknowledgement.---}

We would like to thank N. Yoshinaga and K. Takahashi for useful discussion.
This work was supported by the KAKENHI
Grants No. 23K17668, No. 23K25827, No. 24K00578,
No. JP24K00583, and by the Grant-in-Aid for Transformative
Research Areas (A) “Correlation Design Science”
(KAKENHI Grant No. 25H01249) from JSPS of
Japan.

\bibliography{perturbation}

\appendix

\renewcommand{\appendixname}{SM}

\clearpage

\setcounter{page}{1}
\setcounter{equation}{0}
\setcounter{table}{0}
\setcounter{figure}{0}

\begin{widetext}

\noindent
{\bf Supplementary Material for `Superconductivity-enhanced phonon angular momentum'}
\\ 
N. Okada, P. Werner and S. Hoshino
\\
(\today)

\section{Cartesian vs angular momentum representation}

In the actual calculations, we solve the equations in the Cartesian representation, whereas the spherical-tensor framework is more suitable for a systematic analysis based on angular momentum.
Here, we summarize the correspondence between these two representations \cite{Takahashi26} and the Green's function formalism \cite{Okada26}.
The Eliashberg equations are given in the beginning of SM~C.

\subsection{Electrons}

\subsubsection{Electron Greens' function and self-energy}
We first introduce the electron Green's function.
By using the twelve-component Nambu spinor \cite{Okada26}
\begin{align}
\bm \Psi &= (c_{x\ua},c_{x\da},c_{y\ua},c_{y\da},c_{z\ua},c_{z\da},c^\dg_{x\ua},c^\dg_{x\da},c^\dg_{y\ua},c^\dg_{y\da},c^\dg_{z\ua},c^\dg_{z\da})^\mathrm{T}
     \label{eq:nambu_spinor}
\end{align}
for the electrons in three-fold degenerate $t_{1u}$ orbitals, 
the local Green's function is defined as follows:
\begin{align}
  \check{G}(\tau) 
    &= 
    -\la \mathcal{T}\,\bm\Psi(\tau)\bm\Psi^\dg \ra
    =
    \begin{pmatrix}
        \hat{G}(\tau) & \hat{F}(\tau) \\
        \hat{\bar{F}} (\tau) & \hat{\bar{G}}(\tau)
    \end{pmatrix},  \label{eq:gcheck}
\end{align}
where the site index is omitted by assuming translational symmetry within the solid. 
The $6\times6$ matrix in spin-orbital space is denoted by the hat symbol (~$\hat{\ }~$), while the $12\times12$ matrix 
in Nambu space is denoted by the check symbol (~$\check{\ }$~). 
In Eq.~\eqref{eq:gcheck}, $A(\tau) = \epn^{\tau\mathscr{H}} A  \epn^{-\tau\mathscr{H}}$ is the Heisenberg representation of the operator in imaginary time, and $ \mathcal{T}$ denotes the imaginary-time ordering operator.
The components of the Green's function matrix are explicitly given by
\begin{align}
        \hat{G}_{\gm\sg,\gm'\sg'}(\tau)
        &= - \la \mathcal T c_{\gm\sg}(\tau) c_{\gm'\sg'}^\dg \ra
        \label{eq:hat_G_def},
        \\
        \hat{F}_{\gm\sg, \gm'\sg'}(\tau) 
        &= - \la \mathcal T c_{\gm\sg}(\tau) c_{\gm'\sg'} \ra
        ,
        \\
        \hat{\bar{F}}_{\gm\sg, \gm'\sg'}(\tau) &= - \la \mathcal T c^\dg_{\gm\sg}(\tau) c_{\gm'\sg'}^\dg \ra
        ,\\
        \hat{\bar{G}}_{\gm\sg, \gm'\sg'}(\tau)
        &= - \la \mathcal T c_{\gm\sg}^\dg(\tau) c_{\gm'\sg'} \ra
        .
\end{align}
The local Green's function can be calculated 
once the local self-energy $\check{\Sigma}(\imu\omega_n)$ is given:
\begin{align}
     &\check{G}(\imu\omega_n) 
    = 
    \int_{-D}^{D} \diff\ep 
    \rho(\ep)
    \left[
    \imu\omega_n \check{1} - (\ep - \mu)\check{\tau}_3 - \check{\Sigma}(\imu\omega_n)
    \right]^{-1}, 
    \label{eq:G_loc_def}
\end{align}
with $\check{\tau}_3 = \mathrm{diag}(\hat{1}, -\hat{1})$ and the fermionic Matsubara frequencies $\omega_n = (2n+1)\pi T$ ($n\in\mathbb{Z}$).
The integral over $\ep$ with density of states $\rho(\ep)$ corresponds to the wave-vector integral needed for obtaining the local Green's function.

\subsubsection{Spherical tensor representation of fermionic operators}

The spherical tensor version of the creation operator is given by \cite{Takahashi26}
\begin{align}
    p_{m=-1,\sg}^\dg &\ =\  (c_{\gm=x,\sg}^\dg - \imu c_{\gm=y,\sg}^\dg)/\sqrt 2 ,
    \\
    p_{m=0,\sg}^\dg &\ =\  c_{\gm=z,\sg}^\dg ,
    \\
    p_{m=1,\sg}^\dg &\ =\  - (c_{\gm=x,\sg}^\dg + \imu c_{\gm=y,\sg}^\dg)/\sqrt 2 ,
\end{align}
where $p^\dg_{m\sg}$ denotes the angular-momentum representation and
$c_{\gm\sg}$ the Cartesian representation.
For simplicity, the site index is omitted throughout the following discussion.
Using these operators, one can define various tensor operators such as quadrupole operators as discussed in the main text.

\subsubsection{Electron orbital angular momentum (OAM)}

The orbital angular momentum is defined in terms of the $p$-fermion operators
\begin{align}
    L_{\rm el}^z &= \sum_{m\sg} m p_{m\sg}^\dg p_{m\sg} 
    \\
    &= - \imu \sum_{\sg} (c^\dg _{x\sg} c_{y\sg}
    - c^\dg _{y\sg} c_{x\sg})
    \ =\  \sum_{\gm,\gm' = x,y,z}\sum_\sg c_{\gm\sg}^\dg \lambda^2_{\gm\gm'} c_{\gm'\sg} ,
    \label{eq:OAM_Cart}
\end{align}
where we used the Gall-Mann matrix $\lambda^2_{\gm\gm'}$ in the last expression \cite{Okada26}.
We also have the tensor representation \cite{Takahashi26}
\begin{align}
    L_{\rm el}^z &= -2 \qty[ p^\dg \otimes \tilde p ]_0^{(1)}
    = -2 \sum_{mm'} C_{1m1m'}^{10} \sum_\sg p^\dg_{m\sg} \tilde p_{m'\sg}
\end{align}
with $\tilde p_{m\sg} = (-1)^{1-m} p_{-m,\sg}$,
where the tensor product for electrons has been defined in the main text.

\subsubsection{Pair amplitude}
We consider the anomalous Green's function for fermions, i.e., the pair amplitude
\begin{align}
    F^\dg_{m\sg, m'\sg'} (\tau)
    &\equiv - \la \mathcal T p_{m\sg}^\dg (\tau) p^\dg_{m'\sg'} \ra
    \\
    &=  - F^\dg_{m'\sg', m\sg} (-\tau),
\end{align}
and define the double-tensor representation
\begin{align}
    \qty[ F^\dg (\tau)]_{M_LM_S}^{(L_{\rm C},S_{\rm C})}
    &= \sum_{mm'}\sum_{\sg\sg'} C^{L_{\rm C}M_L}_{1m1m'} C^{S_{\rm C}M_S}_{\frac 1 2 \sg \frac 1 2 \sg'}
        F^\dg_{m\sg, m'\sg'} (\tau).
\end{align}
The static pair amplitude with orbital-rank $L_{\rm C}$ and spin-rank $S_{\rm C}$, which represent the Cooper-pair orbital and spin angular momentum, respectively, is given by
\begin{align}
P_{M_LM_S}^{(LS)} &\equiv - [F^\dg(\tau=0)]_{M_LM_S}^{(L_{\rm C}S_{\rm C})}
= - T \sum_n [F^\dg(\imu\omega_n)]_{M_LM_S}^{(L_{\rm C}S_{\rm C})},
\end{align}
for which $L_{\rm C}+S_{\rm C}=$\,even must be satisfied because of the Pauli exclusion principle, as confirmed by the properties of the Clebsch-Gordan (CG) coefficients \cite{Varshalovich_book}.

\subsection{Phonons}

\subsubsection{Phonon Green's function and self-energy}
\label{sec:phonon_green}
We consider a set of local phonon modes labeled by $\eta$ and introduce the twelve-component spinor \cite{Kaga22,Okada26}
\begin{align}
    \bm{\psi}
    &=
    \bigl(
    \phi_0, \phi_1, \phi_3, \phi_4, \phi_6, \phi_8, 
    \,
    p_0, p_1, p_3, p_4, p_6, p_8
    \bigr)^{\mathrm{T}}. 
    \label{eq:phonon_spinor}
\end{align}
These modes are associated with the creation ($a_\eta^\dg$) and annihilation ($a_\eta$) operators by $\phi_\eta = a_\eta + a_\eta^\dg$ (displacement operator) and $p_\eta = (a_\eta - a_\eta^\dg)/\imu$ (momentum operator).
$\phi_{0,1,3,4,6,8}$ represents the isotropic, $xy$-type, ($x^2-y^2$)-type, $zx$-type, $yz$-type, and ($3z^2-r^2$)-type distortions, respectively.
The canonical commutation relations are
\begin{align}[\phi_{i\eta},p_{j\eta'}]&=2\imu\,\delta_{ij}\delta_{\eta\eta'},
\\
[\phi_{i\eta},\phi_{j\eta'}]&=[p_{i\eta},p_{j\eta'}]=0.
\end{align}
The phonon Green's function is defined in imaginary time as
\begin{align}
    \check{\mathscr G}(\tau)
    &=
    -\langle \mathcal{T}\,\bm{\psi}(\tau)\bm{\psi}^{\mathrm{T}} \rangle
    =
    \begin{pmatrix}
        \hat{\mathscr G}^{\phi\phi}(\tau) & \hat{\mathscr G}^{\phi p}(\tau) \\
        \hat{\mathscr G}^{p\phi}(\tau) & \hat{\mathscr G}^{pp}(\tau)
    \end{pmatrix},
    \label{eq:phonon_Gcheck}
\end{align}
where $\bm{\psi}(\tau)=\epn^{\tau\mathscr{H}}\bm{\psi}\epn^{-\tau\mathscr{H}}$ and
$\mathcal{T}$ denotes the imaginary-time ordering operator.
Here, the hat symbol ($\hat{\ }$) denotes a $6 \times 6$ matrix in the phonon-mode space, whereas the check symbol ($\check{\ }$) denotes a $12 \times 12$ matrix in the $(\phi,p)$ space.
The component representation is explicitly given by
\begin{align}
    \mathscr G^{\phi\phi}_{\eta\eta'}(\tau)
    &= -\langle \mathcal{T}\,\phi_{\eta}(\tau)\phi_{\eta'} \rangle, \label{eq:green_phonon_phiphi}
    \\
    \mathscr G^{\phi p}_{\eta\eta'}(\tau)
    &= -\langle \mathcal{T}\,\phi_{\eta}(\tau)p_{\eta'} \rangle,
    \\
    \mathscr G^{p\phi}_{\eta\eta'}(\tau)
    &= -\langle \mathcal{T}\,p_{\eta}(\tau)\phi_{\eta'} \rangle,
    \\
    \mathscr G^{pp}_{\eta\eta'}(\tau)
    &= -\langle \mathcal{T}\,p_{\eta}(\tau)p_{\eta'} \rangle.
\end{align}

Given the local phonon self-energy $\Pi_{\eta\eta'}^{\phi\phi}(\imu\nu_m)$, the phonon Green's function in Eq.~\eqref{eq:green_phonon_phiphi} can be computed using the Dyson equation 
\begin{align}
    \mathscr G_{\eta\eta'}^{\phi\phi\, -1}(\imu\nu_m) 
    &= 
    \mathscr G_{0, \eta\eta'}^{\phi\phi\, -1}(\imu\nu_m) - \Pi_{\eta\eta'}^{\phi\phi}(\imu\nu_m),
    \label{eq:def_phonon_Green_Matsubara}
    \\
    \mathscr G_{0, \eta\eta'}^{\phi\phi}(\imu\nu_m)
    &=
    \frac{2\omega_\eta}{(\imu\nu_m)^2 - \omega_\eta^2}\delta_{\eta\eta'}, \label{eq:d0}
\end{align}
where $\nu_m = 2m\pi T$ ($m\in\mathbb{Z}$) denotes the bosonic Matsubara frequencies.

To obtain useful relations among these components, we employ the Heisenberg equation of motion in imaginary time,
$\partial_\tau \mathcal{O}(\tau)=[\mathscr{H},\mathcal{O}(\tau)]$.
For the harmonic phonon Hamiltonian, we use
\begin{align}
    \partial_\tau \phi_{\eta}(\tau)
    &=
    -\imu\,\omega_{\eta}\,p_{\eta}(\tau),
    \label{eq:eom_phi}
\end{align}
which is valid even in the presence of electron-phonon interactions.
This leads to the equation of motion for $\mathscr G^{\phi\phi}$,
\begin{align}
    \partial_\tau \mathscr G^{\phi\phi}_{\eta\eta'}(\tau)
    &=
    -\imu\,\omega_{\eta}\,\mathscr G^{p\phi}_{\eta\eta'}(\tau).
    \label{eq:eom_Gphiphi}
\end{align}
Similarly, using $[\phi_{\eta},p_{\eta'}]=2\imu\,\delta_{\eta\eta'}$, we obtain
\begin{align}
    \partial_\tau \mathscr G^{\phi p}_{\eta\eta'}(\tau)
    &=
    -2\imu\,\delta(\tau)\delta_{\eta\eta'}
    -\imu\,\omega_{\eta}\,\mathscr G^{pp}_{\eta\eta'}(\tau).
    \label{eq:eom_Gphip}
\end{align}

We next Fourier transform to bosonic Matsubara frequencies $\nu_m=2\pi m T$ ($m\in\mathbb{Z}$) and set $z=\imu\nu_m$.
With $\partial_\tau \rightarrow -z$, Eqs.~\eqref{eq:eom_Gphiphi} and \eqref{eq:eom_Gphip} become
\begin{align}
    z\,\mathscr G^{\phi\phi}_{\eta\eta'}(z)
    &= 
    \imu\,\omega_{\eta}\,\mathscr G^{p\phi}_{\eta\eta'}(z),
    \label{eq:rel_z_Gphiphi}
    \\
    z^2\,\mathscr G^{\phi \phi}_{\eta\eta'}(z)
    &= 
    2\omega_\eta \delta_{\eta\eta'}
    +
    \omega_{\eta}^2 \,\mathscr G^{pp}_{\eta\eta'}(z).
    \label{eq:rel_z_Gphip}
\end{align}
In addition, from the hermiticity of $\phi_{\eta}$ and $p_{\eta}$, one finds the symmetry relations
\begin{align}
    \mathscr G^{\phi\phi}_{\eta\eta'}(\tau)^{\ast}
    &= \mathscr G^{\phi\phi}_{\eta'\eta}(\tau),
    \qquad
    \mathscr G^{p\phi}_{\eta\eta'}(\tau)^{\ast}
    = \mathscr G^{\phi p}_{\eta'\eta}(\tau),
\end{align}
which are useful for numerical consistency checks.

Finally, introducing the diagonal frequency matrix
$\hat{\omega}=\mathrm{diag}(\omega_0,\omega_1,\omega_1, \omega_1, \omega_1, \omega_1)$,
the relations employed in the actual calculations can be written compactly as
\begin{align}
    z\,\hat{\mathscr G}^{\phi\phi}(z)
    &= \imu\,\hat{\omega}\,\hat{\mathscr G}^{p\phi}(z)
    = -\imu\,\hat{\mathscr G}^{\phi p}(z)\,\hat{\omega},
    \label{eq:matrix_rel_1}
    \\
    z\,\hat{\mathscr G}^{\phi p}(z)
    &= 2\imu\,\hat{1}+\imu\,\hat{\omega}\,\hat{\mathscr G}^{pp}(z).
    \label{eq:matrix_rel_2}
\end{align}
Thus, we only have to calculate $\hat{\mathscr G}_{\phi\phi}$, while the other components are evaluated using the above relations.

\subsubsection{Spherical tensor representation of bosonic operators}

The spherical tensor version of the creation operators is given by
\begin{align}
    s^\dg &\ =\  a_{\eta = 0}^\dg,
    \\
    d_{m=-2}^\dg &\ =\  (a_{\eta=3}^\dg - \imu a_{\eta=1}^\dg)/\sqrt 2,
    \\
    d_{m=-1}^\dg &\ =\  (a_{\eta=4}^\dg - \imu a_{\eta=6}^\dg)/\sqrt 2,
    \\
    d_{m=0}^\dg &\ =\  -a_{\eta=8}^\dg,
    \\
    d_{m=1}^\dg &\ =\  -(a_{\eta=4}^\dg + \imu a_{\eta=6}^\dg)/\sqrt 2,
    \\
    d_{m=2}^\dg &\ =\  (a_{\eta=3}^\dg + \imu a_{\eta=1}^\dg)/\sqrt 2,
\end{align}
where $s^\dg$ is the creation operator for the non-degenerated $A_g$ mode ($\ell=0$) and $d_m^\dg$  for the five-fold degenerate $H_g$ mode ($\ell=2$).

We also define the displacement and momentum tensors by
\begin{align}
    X_m &= d_m^\dg + \tilde d_m ,
    \\
    P_m &= (\tilde d_m - d_m^\dg)/\imu ,
\end{align}
where $m\in [-\ell ,\ell ]$ with $\ell =2$ ($d$-boson), and $    \tilde d_m = (-1)^{\ell-m} d_{-m} = (-1)^m d_{-m}
$.
The canonical commutation relation is given by
\begin{align}
    &[X_m, P_{m'}] = 2\imu (-1)^{m}\delta_{m,-m'}
    \equiv 2\imu I_{mm'},
\end{align}
where $I_{mm'} = (-1)^m \delta_{m,-m'}$.

\subsubsection{Phonon angular momentum (PAM)}
We define the phonon angular momentum in terms of  the $d$-boson operators \cite{Takahashi26} 
\begin{align}
    L_{\mathrm{ph}}^z
    &=
    \sum_mm d_m^\dg d_m
    \\
    &=
    2\imu (-a_3^\dg a_1 + a_1^\dg a_3)
    +
    \imu (-a_4^\dg a_6 + a_6^\dg a_4).
\end{align}
Using $\phi_\eta$ and $p_\eta$, we obtain 
\begin{align}
    L_{\mathrm{ph}}^z
    &=
    \phi_3 p_1 - p_3\phi_1 
    +
    \frac{1}{2} (\phi_4 p_6 - p_4\phi_6),
\end{align}
which is analogous to the standard angular momentum $L^z = (\bm r\times \bm p)_z = xp_y -y p_x$.
There also exists another tensor expression:
\begin{align}
    L_{\rm ph}^z &= \sqrt{10}\  \qty[d^\dg \otimes\tilde d]_0^{(1)} = \sqrt{10}\sum_{mm'} C^{10}_{2m2m'} d_m^\dg \tilde d_{m'}
    \\
    &= \imu \sqrt{\frac 5 2} \  \qty[X\otimes P]_0^{(1)},
    \label{eq:def_of_Lz_XP}
\end{align}
where the tensor product has been defined in the main text.

\subsubsection{Tensorial description of Green's functions and self-energies}

The phonon Green's function in the angular momentum basis is given by
\begin{align}
    D_{mm'} (\tau) &= - \la \mathcal T X_{m} (\tau) X_{m'} \ra,
    \\
    D_{0,mm'} (\tau) &=  D_0(\tau) I_{mm'},
\end{align}
where $D_0$ is the non-interacting Green's function.
The Dyson equation in the Fourier domain reads 
\begin{align}
    D_{mm'}(\imu\nu) &= D_{0,mm'}(\imu\nu) + \big[ D_{0}(\imu\nu)\cdot \Pi (\imu\nu) \cdot D(\imu\nu) \big]_{mm'}
    \\
&= D_{0,mm'}(\imu\nu) + \sum_{m_1m_2}(-1)^{m_1+m_2}D_{0,mm_1}(\imu\nu) \Pi_{-m_1,-m_2}(\imu\nu) D_{m_2 m'}(\imu\nu)
.
\end{align}
Now we define the spherical-tensor representation
\begin{align}
    D_M^{(L)}(\imu \nu) &= \sum_{mm'} C^{LM}_{2 m2 m'} D_{mm'}(\imu \nu)
\end{align}
 and its inverse
\begin{align}
    D_{mm'}(\imu \nu) &= \sum_{L=0}^4 \sum_{M=-L}^L C^{LM}_{2 m 2 m'} D_M^{(L)}(\imu \nu).
\end{align}
A similar relation is applied to the phonon self-energy $\Pi$.
Then
\begin{align}
    D_M^{(L)} &= D_{0,M}^{(L)}
    + \sum_{L_1L_2L_3}\sum_{L'} (-1)^{L_2+L_3+L'+L} \sqrt{\frac{2L'+1}{2L+1}} 
    \left\{
        \begin{matrix}
            2&2&L_2\\
            2&2&L\\
            L_3&L_1&L'
        \end{matrix}
    \right\}
    \qty[ \qty[D_0^{(L_1)}\otimes D^{(L_3)}]^{(L')}\otimes \Pi^{(L_2)} ]^{(L)}_M
    \nonumber \\
    &\hspace{30mm} \times \sqrt{(2L_1+1)(2L_2+1)(2L_3+1)(2L+1)},
\end{align}
where we have used the formula in Ref.~\cite{Varshalovich_book} for a four CG-coefficient product, and $\{\cdots \}$ is the Wigner $9j$ symbol.
We have also defined
\begin{align}
    &D_{0,M}^{(L)} = \sqrt{2\ell+1} \ \delta_{L0}\delta_{M0}D_0 ,
    \\
    &\qty[A^{(L_1)} \otimes B^{(L_2)}]_{M}^{(L)}
    = \sum_{M_1=-L_1}^{L_1} \sum_{M_2=-L_2}^{L_2} C^{LM}_{L_1M_1L_2M_2} A_{M_1}^{(L_1)}  B_{M_2}^{(L_2)} .
\end{align}
Specifically, considering the fact that $D_0$ has only an $L=0$ component, we find 
\begin{align}
    D_M^{(L)}(\imu\nu) &= 
    \sqrt{2\ell + 1}\, D_0 (\imu\nu)
    \qty( \delta_{L0}
    \delta_{M0} + \sum_{L_1L_2} S_{L_1L_2}^L \qty[    \Pi^{(L_1)}(\imu\nu)\otimes D^{(L_2)} (\imu\nu) ]^{(L)}_M  ),
    \\[2mm]
    S_{L_1L_2}^L &= (-1)^L  \sqrt{\frac{(2L_1+1)(2L_2+1)}{2\ell + 1}}
    \left\{
        \begin{matrix}
            L_1 &L_2 &L\\
            \ell&\ell&\ell 
        \end{matrix}
    \right\},
\end{align}
where we have used the following reduction from the $9j$ to the $6j$ symbol \cite{Varshalovich_book}:
\begin{align}
     \left\{
        \begin{matrix}
            2&2&L_2\\
            2&2&L\\
            L_3&0&L_3
        \end{matrix}
    \right\}
    &= \frac{(-1)^{L+L_3}}{\sqrt{5(2L_3+1)}} 
    \left\{
        \begin{matrix}
            2&L&2\\
            L_2&2&L_3
        \end{matrix}
    \right\}.
\end{align}
For a small self-energy, we obtain
\begin{align}
    D_M^{(L)}(\imu\nu) &\simeq 
\sqrt{5}\, D_0 (\imu\nu)
    \delta_{L0}
    \delta_{M0} 
    +
    D_0 (\imu\nu)^2
      \Pi^{(L)}_M(\imu\nu).
\end{align}
Note that in the above expressions the specific condition $\ell=2$ has been used.

Relations among the Green's functions similar to those in Sec.~\ref{sec:phonon_green} can be derived.
We introduce 
\begin{align}
    D_{mm'}^{XX}(\tau) &= - \la \mathcal T X_m(\tau) X_{m'} \ra = D_{mm'}(\tau),
    \\
    D_{mm'}^{PP}(\tau) &= - \la \mathcal T P_m(\tau) P_{m'} \ra,
    \\
    D_{mm'}^{XP}(\tau) &= - \la \mathcal T X_m(\tau) P_{m'} \ra,
    \\
    D_{mm'}^{PX}(\tau) &= - \la \mathcal T P_m(\tau) X_{m'} \ra.
\end{align}
The equation of motion gives
\begin{align}
    - \partial_\tau D^{XX}(\tau) &=  \imu\omega_1 D^{PX}(\tau),
    \\
    -\partial_\tau D^{XP}(\tau) &= 2\imu I \delta(\tau) + \imu \omega_1 D^{PP}(\tau). 
\end{align}
Using the relation
\begin{align}
    D^{AB}_{mm'}(\tau) &=  D^{BA}_{m'm} (-\tau),
    \\
    D^{AB}_{mm'}(\tau)^* &=  (-1)^{m+m'} D^{BA}_{-m',-m} (\tau),
\end{align}
where $A,B = X,P$, we obtain
\begin{align}
    D_{mm'}^{XP}(\tau) &= - D_{mm'}^{PX}(\tau).
\end{align}
The angular momentum defined in Eq.~\eqref{eq:def_of_Lz_XP} is evaluated from $D^{XP}$, which can in turn be obtained from $D^{XX}$ by using the above relations; therefore, only the latter is required.

We use the following relation between the Cartesian and tensor representations:
\begin{align}
X_m = 
    \begin{pmatrix}
        X_{-2} \\ X_{-1} \\ X_{0} \\ X_{1} \\ X_{2}
    \end{pmatrix}
    &= \frac{1}{\sqrt 2}
    \begin{pmatrix}
        -\imu&1&0&0&0\\
        0&0&1&-\imu&0\\
        0&0&0&0&-\sqrt{2}\\
        0&0&1&\imu&0\\
        \imu&1&0&0&0
    \end{pmatrix}
    \begin{pmatrix}
        \phi_1 \\ \phi_3 \\ \phi_4 \\ \phi_6 \\ \phi_8
    \end{pmatrix}
    = \sum_\eta U_{m\eta} \phi_\eta,
\end{align}
where $U$ satisfies $U^{-1}=U^\dg$.
With this, one can write
\begin{align}
    \hat {\mathscr G}^{\phi\phi}(\tau) &= \hat U^\dg \hat D^{XX} (\tau) \hat U^*,    
\end{align}
where we only considered the anisotropic components.

\section{Semiclassical equation of motion for the phonons}

Given the phonon Green's function, the effective one-body action for the phonons can be written as
\begin{align}
    S 
    &= - \sum_{m} \sum_{\eta\eta'} 
    \phi_\eta (-\imu\nu_m)
    (\mathscr G^{\phi\phi})^{-1}_{\eta\eta'} (\imu\nu_m)
    \phi_{\eta'}(\imu\nu_m).
\end{align}
Indeed, one can verify that 
$    \mathscr G^{\phi\phi}_{\eta\eta'}(\imu \nu_m) = - \la \phi_\eta(\imu\nu_m) \phi_{\eta'}(-\imu\nu_m) \ra
$,
where the expectation value is defined by
$ 
\displaystyle \la \cdots \ra 
    =
    \frac{\int D\phi \, (\cdots)\epn^{-S}}
    {\int D\phi \, \epn^{-S}}
$.
In the low-energy regime, retaining the low-frequency contributions and terms linear in the external field, the phonon action for the Cartesian components $\eta = 4,6$ (corresponding to $zx$- and $yz$-type distortions) is approximately given by
\begin{align}
    S 
    &\simeq
    -\sum_{m}\sum_{\eta,\eta'=4,6}
    \phi_\eta (-\imu\nu_m)
    \left[
    \left(
    \frac{(\imu\nu_m)^2 - \omega_1^2}{2\omega_1}
    -\Pi_0 
    \right)
    \delta_{\eta\eta'}
    +
    \Gamma \epsilon_{\eta\eta'}\nu_m
    \right]
    \phi_{\eta'}(\imu\nu_m). 
\end{align}
Here we used $\Pi_{64}(\imu\nu_m) \simeq \Gamma \nu_m$ and introduced the two-dimensional antisymmetric tensor
$\epsilon$.
The constant $\Pi_{0}$ denotes the low-frequency limit of the diagonal component of the phonon self-energy.

Taking the variation of the action,
$\delta S / \delta\phi_\eta = 0$,
we obtain
\begin{align}
\begin{cases}
    \displaystyle
    \left(
    \frac{(\imu\nu_m)^2 - \omega_1^2}{2\omega_1} - \Pi_0 
    \right)\phi_4
    +
    \Gamma \nu_m\phi_6
    &=
    0,
    \\[5mm]
    \displaystyle
    \left(
    \frac{(\imu\nu_m)^2 - \omega_1^2}{2\omega_1} - \Pi_0 
    \right)\phi_6
    -
    \Gamma\nu_m\phi_4
    &=
    0.
\end{cases}
\end{align}

By performing the analytic continuation and Fourier transformation,
$\imu\nu_m \rightarrow \omega \rightarrow \imu\partial_t$,
we obtain the equation of motion in the real-time domain:
\begin{align}
\begin{cases}
    -\ddot{\phi}_4 - \tilde{\omega}_1^2\phi_4 - 2\Gamma\omega_1\dot{\phi}_6
    &= 0,
    \\[3mm]
     -\ddot{\phi}_6 - \tilde{\omega}_1^2\phi_6 - 2\Gamma\omega_1\dot{\phi}_4
    &= 0,
\end{cases}
\label{eq:eom}
\end{align}
where $
    \tilde{\omega}_1^2 = \omega_1^2 + 2\omega_1\Pi_0
$ is the renormalized phonon frequency.
Physically, the third term with $\Gamma$ acts as a velocity-coupling term in the $(\phi_4,\phi_6)$ space and breaks time-reversal symmetry.
Consequently, this term bends the trajectory of the phonon without dissipation, in a manner analogous to the Lorentz force.
As a result, circularly polarized phonons emerge through the hybridization of the linearly polarized modes $\phi_{4,6}$.

Equation~\eqref{eq:eom} constitutes a set of coupled differential equations in which $\phi_4$ and $\phi_6$ are coupled through $\Gamma$.
To decouple these equations and clarify the eigenmodes, we introduce the circularly polarized basis
\begin{align}
    \phi_\pm = \phi_4 \pm \imu\phi_6.
\end{align}
Diagonalizing the equations of motion in this basis yields the eigenfrequencies
\begin{align}
    \Omega_\pm
    &\simeq
     \tilde{\omega}_1  \pm \omega_1\Gamma.
\end{align}
Here we assumed that the effective magnetic field $\Gamma$ is sufficiently small compared to the phonon frequency and retained only the linear contribution in $\Gamma$.
The eigenfrequencies exhibit a Zeeman-like splitting of magnitude $\pm \omega_1\Gamma$ between the positive- and negative-circular-polarization modes induced by the effective magnetic field.
The corresponding eigenvectors are given by
$
    \bm{v}_\pm \propto (1, \pm\imu)^{\mathrm{T}},
$
which describe circular motion in the $(\phi_4,\phi_6)$ plane.

\section{Low-energy effective theory}

\subsection{Perturbation theory for the phonon self-energy}
The phonon self-energy in the Cartesian basis is given by \cite{Okada26}
\begin{align}
    \Pi_{\eta\eta'}(\tau) &=
    \frac 1 2
    g_{\eta}g_{\eta'}  
    \trace\,
    \Bigl[
        \check{G}(- \tau)
        \check \lambda^{\eta'}
        \check{G}(\tau)
        \check \lambda^{\eta}
    \Bigr], 
\end{align}
where we consider the anistropic modes with $\eta = 1,3,4,6,8$.
The phonon Green's function is given by
\begin{align}
    D_{\eta\eta'}^{-1}(\imu\nu_m) &= D_{0, \eta\eta'}^{-1}(\imu\nu_m) - \Pi_{\eta\eta'}(\imu\nu_m),
    \\
    D_{0, \eta\eta'}(\imu\nu_m)
    &=
    \frac{2\omega_\eta}{(\imu\nu_m)^2 - \omega_\eta^2}\delta_{\eta\eta'}.
\end{align}
In this section we use the notation $D_{\eta\eta'} = \mathscr G^{\phi\phi}_{\eta\eta'}$.

Now we consider the linear response with respect to the external field $h = \mathrm g_L \mu_{\rm B}B$ applied along $z$ direction:
\begin{align}
    \Pi_{1,\eta\eta'}(\tau)
    &= \frac 1 2 g_1^2 {\rm tr\,} \qty[\check G_0(-\tau)\check \lambda^{\eta'} \check G_1(\tau)\check \lambda^\eta
    + \check G_1(-\tau)\check \lambda^{\eta'} \check G_0(\tau)\check \lambda^\eta],
\end{align}
where $\check \lambda^\eta = \hat \lambda^\eta \otimes 1_{\rm spin} \otimes \tau^3_{\rm Nambu}$, with $\hat \lambda^\eta$ denoting the Gell-Mann matrices, and the Kronecker products taken with the identity matrix in spin space and the matrix $\tau^3 = {\rm diag}(1,-1)$ in Nambu space \cite{Okada26}.
The perturbed phonon Green's function is given by
\begin{align}
D_{\eta\eta'} (\imu\nu_m) &= 
    D_0(\imu\nu_m) \delta_{\eta\eta'}
    + D_{1,\eta\eta'}(\imu\nu_m),
    \\
D_{1,\eta\eta'} (\imu\nu_m) &=     
D_{0}(\imu\nu_m) ^2
\Pi_{1,\eta\eta'} (\imu\nu_m),
\end{align}
and the electron Green's function by \cite{Okada26}
\begin{align}
    \check G_0(\imu\omega_n)
     &= \begin{pmatrix}
         G_0(\imu\omega_n) \hat 1 &  F_0(\imu\omega_n) \hat \epsilon \\
         F_0(\imu\omega_n) \hat \epsilon^{\rm T} &
         G_0(\imu\omega_n)\hat 1
     \end{pmatrix},
     \\
    \check G_1(\imu\omega_n)
     &= \begin{pmatrix}
         G_1(\imu\omega_n) \hat \lambda^2 &  F_1(\imu\omega_n) \hat \lambda^2\hat \epsilon \\
         F_1(\imu\omega_n) \hat \epsilon^{\rm T} \hat \lambda^2 &
         - G_1(\imu\omega_n)\hat \lambda^{2\rm T}
     \end{pmatrix},
\end{align}
where $(\hat 1)_{\gm\sg,\gm'\sg'} = \delta_{\gm\gm'}\delta_{\sg\sg'}$ and 
$(\hat\epsilon) _{\gm\sg,\gm'\sg'} = \delta_{\gm\gm'}\epsilon_{\sg\sg'}$ with $\epsilon_{\sg\sg'}$ being the two-dimensional antisymmetric tensor.
The Gell-Mann matrix $\hat \lambda^2$ arises from the angular momentum along the $z$ direction in Eq.~\eqref{eq:OAM_Cart}.
The explicit forms of $G_{0,1}$ and $F_{0,1}$ are considered later.
The phonon self-energy may thus be written as
\begin{align}
        \Pi_{1,\eta\eta'}(\tau)
    &= - g_\eta g_{\eta'}
    \big[ G_0(\tau)G_1(\tau) - F_0(\tau) F_1(\tau)\big]
    \ 
    {\rm tr\,}\hat \lambda^2 [\hat \lambda^{\eta},\hat \lambda^{\eta'}  ].
\end{align}
The commutator becomes nonzero only for
\begin{align}
    [\hat \lambda^1,\hat\lambda^3] &= -2\imu \hat \lambda^2,
    \\
    [\hat \lambda^4,\hat\lambda^6] &= \imu \hat \lambda^2.
\end{align}
Let us parametrize the self-energy as
\begin{align}
    \Pi^{(1)}_{\eta\eta'} &= \left\{ 
    \begin{matrix}
     \Pi_1 & (\eta=4,\eta'=6)   
     \\[2mm]
     - \Pi_1 & (\eta=6,\eta'=4)   
     \\[2mm]
     - 2 \Pi_1 & (\eta=1,\eta'=3)   
     \\[2mm]
     2 \Pi_1 & (\eta=3,\eta'=1)   
    \end{matrix}
    \label{eq:Pi_induced_by_h}
    \right.,
    \\[2mm]
    \Pi_1(\tau) &= - 4 \, \imu \, g_1^2 \big[ G_0(\tau)G_1(\tau) - F_0(\tau) F_1(\tau)\big].
\end{align}
The Fourier transform gives
\begin{align}
        \Pi_1(\imu \nu_m) 
        &= - 4\imu g_1^2 T\sum_n [G_0(\imu\omega_n + \imu \nu_m) G_1(-\imu\omega_n) - F_0(\imu\omega_n + \imu \nu_m) F_1(-\imu\omega_n)].
\end{align}

\subsection{Phonon self-energy in the superconducting state}

Now we consider the superconducting state at sufficiently low temperature (without Fermi surface).
We are interested in the small bosonic frequency behavior, and expand the self-energy with respect to $\nu_m$ as
\begin{align}
        \Pi_1(\imu \nu_m) 
        &= 4 \nu_m g_1^2 T\sum_n [G_0'(\imu\omega_n) G_1(-\imu\omega_n) - F_0'(\imu\omega_n) F_1(-\imu\omega_n)]
        + O(\nu_m^3).
\end{align}
In the evaluation of the phonon self-energy, we neglect the electron self-energy except for the static pairing potential.
The zeroth-order Green's function is given by
\begin{align}
    G_0(\imu\omega_n)
    &= \int \diff \ep \rho(\ep) \frac{\imu\omega_n + \ep}{(\imu\omega_n)^2 - \ep^2 - \Delta^2}
    \\
     &= - \frac{\imu (\imu\omega_n)}{\Omega_n} g(\imu\Omega_n),
     \\
     g(\imu\Omega_n) &= \frac{2}{D^2} \qty( \imu\Omega_n - \imu \sqrt{D^2+\Omega_n^2} ),
\end{align}
where the pair potential $\Delta \in \mathbb R$ is assumed to be constant (BCS limit, i.e., no retardation effect), and we have defined $\Omega_n = \sqrt{\omega_n^2+\Delta^2}$.
The derivative is given by 
\begin{align}
    G_0'(z) &= - \int \diff \ep \rho(\ep)  \frac{z^2+\ep^2 + \Delta^2}{(z^2-\ep^2-\Delta^2)^2}
    \\
    &= - \int \diff \ep \rho(\ep)  \frac{(z+\ep)^2 + \Delta^2}{(z^2-\ep^2-\Delta^2)^2}
    \\
    &= - \int \diff \ep \rho(\ep) (g^2+f^2)
    \\
    &= - G_1(z)/h ,
\end{align}
where $z=\imu\omega_n$.
The derivative evaluates to
\begin{align}
    G_0'(z) &= g'(\imu\Omega) + \qty( - g'(\imu\Omega) + \frac{g(\imu\Omega)}{\imu\Omega} ) \frac{\Delta^2}{\Omega^2},
\end{align}
where
\begin{align}
    g'(z) &=\frac{2}{D^2} \qty(1 + \frac{\imu z}{\sqrt{D^2-z^2}}),
\end{align}
for ${\rm Im\,}z>0$.
The anomalous function is given by
\begin{align}
    F_0(z) &= \int \diff \ep \rho(\ep) \frac{\Delta}{z^2-\ep^2 -\Delta^2}
    = \Delta \frac{g(\imu\Omega)}{\imu\Omega}.
\end{align}
The derivative is 
\begin{align}
    F_0'(z) &= - \int \diff \ep \rho(\ep) \frac{2z\Delta}{(z^2-\ep^2 -\Delta^2)^2}
    \\
    &= - \int \diff \ep \rho(\ep) (g+\bar g) f
    = \qty( - g'(\imu\Omega) + \frac{g(\imu\Omega)}{\imu\Omega} )  \frac{z\Delta}{\Omega^2}
    \\
    &= - F_1(z)/h .
\end{align}
The phonon self-energy is then given by
\begin{align}
    \Pi_1 (\imu \nu_m) &= -4h\nu_m g_1^2 T \sum_n [G_0'(\imu\omega_n)G_0'(-\imu\omega_n) - F_0'(\imu\omega_n) F_0'(-\imu\omega_n)]
    \\
    &\simeq -4h\nu_m g_1^2 T \sum_n \qty(- \frac{2}{D\Omega})^2 \frac{\Delta^4 - \omega_n^2\Delta^2}{\Omega^4}
    \\
    &= -\frac{16 h \nu_m g_1^2 \Delta^2}{D^2} T\sum_n \frac{\Delta^2 - \omega_n^2}{\Omega^6}
    \\
    &\simeq -\frac{2 h \nu_m g_1^2}{D^2 \Delta} ,
\end{align}
where $T\ll \Delta \ll D$ has been used.

\subsection{Phonon self-energy in the normal state}

\subsubsection{Analysis with the Wavenumber sum performed first}
The self-energy in the normal state (with Fermi surface) is given by
\begin{align}
    \Pi_1(\imu\nu_m) &= 4\imu h g_1^2 T\sum_n G_0(\imu \omega_n + \imu \nu_m) G_0'(-\imu\omega_n),
    \\
    G_0(\imu\omega_n) &= \int \diff \ep \frac{\rho(\ep)}{z-\ep} = \frac{2}{D^2} \qty( \imu\omega_n - \imu {\rm sgn\,}\omega_n\, \sqrt{D^2+\omega_n^2} ).
    \label{eq:G0_expr}
\end{align}
We consider the coefficient of the $\nu_m$-linear part, for which we assume $\nu_m \neq 0$.
The derivative is given by
\begin{align}
    G_0'(\imu\omega_n) = \frac{1}{\imu} \frac{\partial G_0(\imu\omega_n)}{\partial \omega_n}
    &= \frac{2}{D^2}  \qty( 1 - \frac{|\omega_n|}{\sqrt{\omega_n^2+D^2}} - 2D \delta(\omega_n) )
    \\
    &\equiv G_{0,{\rm reg}}(\imu\omega_n) + G_{0,{\rm sing}}(\imu\omega_n),
\end{align}
where $G_{0,{\rm reg}}(\imu\omega_n) > 0$ and $G_{0,{\rm sing}}(\imu\omega_n) < 0$.
Note that delta function is not needed for $\omega_n \neq 0$, but for the $\nu_m$ expansion its contribution is nontrivial because we regard the Green's function as an analytic function of a continuous variable.
Namely, the correct low-$T$ limit is taken first by sending $T\to 0$ and writing
\begin{align}
    \Pi_1(\imu \nu) &= -
    4 h g_1^2 \nu \int_{-\infty}^\infty \frac{\diff \omega}{2\pi} \Big[ G'_{0,\rm reg}(\imu \omega) + G'_{0,\rm sing}(\imu \omega) \Big] G'_{0,\rm reg}(-\imu \omega)
    \\
    &= -\frac{4h\nu g_1^2}{2\pi}
    \frac{4}{D^4}
    \int_{-\infty}^{\infty} \diff \omega
    \qty( 1 - \frac{|\omega_n|}{\sqrt{\omega_n^2+D^2}} - 2D \delta(\omega_n) )\qty( 1 - \frac{|\omega_n|}{\sqrt{\omega_n^2+D^2}} ).
\end{align}
The validity is confirmed by comparing it with the numerical result.
It is clear from this expression that the delta function (singular) contribution has a different sign compared to the other (regular) contributions.
We estimate this singular contribution from delta function as
\begin{align}
    \Pi_{1,\rm sing}(\imu \nu)
    &= \frac{16 h\nu g_1^2}{\pi D^3}.
\end{align}
The others yield
\begin{align}
    \Pi_{1,\rm reg}(\imu \nu) &= -\frac{4h\nu g_1^2}{2\pi}
    \frac{4}{D^4}
    \int_{-\infty}^{\infty} \diff \omega
    \qty( 1 - \frac{|\omega_n|}{\sqrt{\omega_n^2+D^2}} ) ^2
    \\
    &= -\frac{4h\nu g_1^2}{2\pi}
    \frac{4}{D^3} (4-\pi).
\end{align}
The total contribution is then obtained as 
\begin{align}
    \Pi_1(\imu \nu) &= \frac{16 h \nu g_1^2}{2\pi D^3} (\pi-2),
\end{align}
where the singular contribution is dominant.
We note that the relation $h \ll T \ll \nu_m$ among the energy scales are assumed for the normal state.

The sign of the singular part is different from the regular part.
This conclusion is obtained for a semi-circular density of states of the conduction electrons, 
$\displaystyle \rho(\ep) = \frac{2}{\pi D^2}\sqrt{D^2-\ep^2}$,
but is qualitatively unchanged if we consider the trapezoidal shape
$\displaystyle \rho(\ep) = \frac{\rho_2 - \rho_1}{2D} \ep + \frac{1}{2D}$ [with the half bandwidth $D = 1/(\rho_1 + \rho_2)$]
or Lorentzian shape
$\displaystyle \rho(\ep) = \frac 1 \pi \frac{D}{(\ep-E_0)^2+D^2}$.
Hence, we expect it to be a generic feature that the sign of the singular [or Fermi-surface (see the next subsection)] contribution is opposite to that of the remaining (regular or Fermi-volume) contribution.
The relative magnitudes, however, may depend on the specific details of the electronic band structure.
In general, the singular contribution should dominate when the density of states at the Fermi level is sufficiently large.

\subsubsection{Analysis with the Matsubara sum performed first}

We show that the singular contribution above originates from the presence of the Fermi surface.
Let us consider the self-energy in a different way.
We begin with the self-energy expressed in terms of the spectral representation of the electron Green's function:
\begin{align}
    \Pi_1(\imu\nu_m) &= 4\imu h g_1^2 T\sum_n \int \diff \ep \diff \ep' \rho(\ep)\rho(\ep') \frac{1}{(\imu\omega_n + \imu\nu_m - \ep)(\imu\omega_n - \ep')^2}.
\end{align}
Here, the $\ep$-integral originates from the wavenumber summation.
The Matsubara sum can be performed using a partial fraction decomposition:
\begin{align}
    \Pi_1(\imu\nu_m) &= 4\imu h g_1^2 \int \diff \ep \diff \ep' \rho(\ep)\rho(\ep') \qty(\frac{f(\ep) - f(\ep')}{(\ep' + \imu\nu_m - \ep)^2} + \frac{f'(\ep')}{\ep' + \imu\nu_m - \ep} ).
\end{align}
The second term gives the Fermi surface contribution at low $T$:
\begin{align}
    \Pi_{1,\rm FS}(\imu\nu_m) &= -4\rho(0) \imu h g_1^2 G_0(\imu \nu_m)
    \\
    &\simeq - 4 \frac{2}{\pi D} \imu h g_1^2 \Big[ G_0(0) + \imu \nu_m \frac{2}{D^2} \Big],
\end{align}
where we have used Eq.~\eqref{eq:G0_expr} for a small $\nu_m$ ($\neq 0$).
The $\nu_m$-linear term is identical to $\Pi_{1,\rm sing}$. 
Hence, the delta function (singular) contribution is indeed originating from the Fermi surface.

\subsection{Evaluation of the phonon angular momentum}

Finally, we evaluate the angular momentum:
\begin{align}
    \la L_{\rm ph}^z\ra &= \frac{\imu c}{2} \left\la \qty[X\otimes P]_0^{(1)} \right\ra
    \\
    &= - \frac{\imu \sqrt{10}}{2} T\sum_m \qty[D^{XP}(\imu \nu_m)]_1^{(1)}
    \\
    &= \frac{ \sqrt{10}}{2\omega_1} T\sum_m \imu\nu_m \qty[D^{XX}(\imu \nu_m)]_0^{(1)}
    \\
    &\simeq \frac{ \sqrt{10}}{2\omega_1} T\sum_m \imu\nu_m \qty[ \Pi(\imu\nu_m)]_{0}^{(1)} D_0(\imu\nu_m)^2
    \\
    &\equiv - \sqrt{10}\ \Gamma \ \mathcal I_T .
\end{align}

Let us consider the rank-1 component of the self-energy in the form
\begin{align}
    [\Pi(\imu\nu)]_{M=0}^{(L=1)} &= \frac{1}{\sqrt{10}} \Big[2\imu (\Pi_{13}-\Pi_{31}) + \imu (\Pi_{46} - \Pi_{64}) \Big]
    \\
    &= - \imu \sqrt{\frac{18}{5}}\ \Pi_1(\imu\nu).
\end{align}
Inspired by the numerical results for the self-energy presented in the main text, we approximate the self-energy as
\begin{align}
    \Pi_1(\imu\nu_m)  &=  \left\{
    \begin{matrix}
        \Gamma \ \nu_m & (|\nu_m| < \ep_c) 
        \\[2mm]
        \Gamma \ep_c \  {\rm sgn\,}\nu_m & (|\nu_m| > \ep_c) \\
    \end{matrix}    \right. ,
\end{align}
where the characteristic energy is $\ep_c \sim \Delta $ for the superconducting (S) state and $\ep_c \sim D$ for the normal (N) state.
Note that the self-energy for $|\nu_m|\lesssim \omega_1$ dominantly contributes to PAM.
Then, at low temperature, 
\begin{align}
    \mathcal I_{T\to 0} (\omega_1,\ep_c) &=  4\omega_1 \int_0^{\ep_c} \diff \nu \frac{\nu^2}{(\nu^2+\omega_1^2)^2}
    +4\omega_1 \ep_c \int_{\ep_c}^{\infty} \diff \nu  \frac{\nu}{(\nu^2+\omega_1^2)^2}
    \\
    &= \left\{
    \begin{matrix}
        2 \ep_c /\omega_1 & & (\ep_c \ll \omega_1) 
        \\[2mm]
        \pi & & (\ep_c \gg \omega_1)
    \end{matrix}
     \right. ,
\end{align}
where we assumed $\Delta \ll \omega_1 \ll D$.
Hence
\begin{align}
    \frac{\la L_{\rm ph}^z \ra_{\rm S}}{\la L_{\rm ph}^z \ra_{\rm N}}
    &= \frac{\Gamma_{\rm S}}{\Gamma_{\rm N}}\cdot \frac{2\ep_{c,{\rm S}}}{\pi \omega_0}
    \\
    &\sim - \frac{D}{\Delta} \cdot \frac{\Delta}{\omega_0}
    = - \frac{D}{\omega_0} ,
\end{align}
which is qualitatively consistent with the numerical calculations.

\vspace{5mm}
\section*{References}

See the list of references in the main text.

\end{widetext}

\end{document}